\documentclass[iop]{emulateapj}
\usepackage{graphics}
\usepackage{amsmath}

\newcommand{\ltsimeq}{\la}
\newcommand{\gtsimeq}{\ga}

\newcommand{\msun}{M$_{\odot}$}

\newcommand{\HI}{H{\sc i}}
\newcommand{\HII}{H{\sc ii}}

\accepted{January 2015}
\shortauthors{McQuinn et al.}
\shorttitle{Characterizing the Star Formation of the low-mass SHIELD Galaxies}

\begin{document}
\title{Characterizing the Star Formation of the Low-Mass SHIELD Galaxies from $\textit{Hubble Space Telescope}$ Imaging\protect\footnotemark[*]}
\footnotetext[*]{Based on observations made with the NASA/ESA Hubble Space Telescope, obtained from the Data Archive at the Space Telescope Science Institute, which is operated by the Association of Universities for Research in Astronomy, Inc., under NASA contract NAS 5-26555.}
\author{Kristen.~B.~W. McQuinn\altaffilmark{1}, 
John M.~Cannon\altaffilmark{2},
Andrew E.~Dolphin\altaffilmark{3},
Evan D. Skillman\altaffilmark{1},
Martha P.~Haynes\altaffilmark{4},
Jacob E.~Simones\altaffilmark{1},
John J.~Salzer\altaffilmark{5},
Elizabeth A.~K.~Adams\altaffilmark{6},
Ed C.~Elson\altaffilmark{7},
Riccardo Giovanelli\altaffilmark{4},
J\"{u}rgen Ott\altaffilmark{8}
}

\altaffiltext{1}{Minnesota Institute for Astrophysics, School of Physics and
Astronomy, 116 Church Street, S.E., University of Minnesota,
Minneapolis, MN 55455, \ {\it kmcquinn@astro.umn.edu}} 
\altaffiltext{2}{Department of Physics and Astronomy, 
Macalester College, 1600 Grand Avenue, Saint Paul, MN 55105, USA}
\altaffiltext{3}{Raytheon Company, 1151 E. Hermans Road, Tucson, AZ 85756, USA}
\altaffiltext{4}{Center for Radiophysics and Space Research, Space Sciences Building, Cornell University, Ithaca, NY 14853, USA}
\altaffiltext{5}{Department of Astronomy, Indiana University, 727 East 3rd Street, Bloomington, IN 47405, USA}
\altaffiltext{6}{Netherlands Institute for Radio Astronomy (ASTRON), Postbus 2, 7900 AA Dwingeloo, The Netherlands}
\altaffiltext{7}{Astrophysics, Cosmology and Gravity Centre (ACGC), Department of Astronomy, University of Cape Town, Private Bag X3, Rondebosch 7701, South Africa}
\altaffiltext{8}{National Radio Astronomy Observatory, P.O. Box O, 1003 Lopezville Road, Socorro, NM 87801, USA}

\begin{abstract}
The Survey of \HI\ in Extremely Low-mass Dwarfs (SHIELD) is an on-going multi-wavelength program to characterize the gas, star formation, and evolution in gas-rich, very low-mass galaxies that populate the faint end of the galaxy luminosity function. The galaxies were selected from the first $\sim10$\% of the \HI\ ALFALFA survey based on their low \HI\ mass and low baryonic mass. Here, we measure the star-formation properties from optically resolved stellar populations for 12 galaxies using a color-magnitude diagram fitting technique. We derive lifetime average star-formation rates (SFRs), recent SFRs, stellar masses, and gas fractions. Overall, the recent SFRs are comparable to the lifetime SFRs with mean birthrate parameter of 1.4, with a surprisingly narrow standard deviation of 0.7. Two galaxies are classified as dwarf transition galaxies (dTrans). These dTrans systems have star-formation and gas properties consistent with the rest of the sample, in agreement with previous results that some dTrans galaxies may simply be low-luminosity dIrrs. We do not find a correlation between the recent star-formation activity and the distance to the nearest neighboring galaxy, suggesting that the star-formation process is not driven by gravitational interactions, but regulated internally. Further, we find a broadening in the star-formation and gas properties (i.e., specific SFRs, stellar masses, and gas fractions) compared to the generally tight correlation found in more massive galaxies. Overall, the star-formation and gas properties indicate these very low-mass galaxies host a fluctuating, non-deterministic, and inefficient star-formation process.
\end{abstract} 


\section{Introduction\label{intro}}
The evolution of a galaxy is often strongly affected by the star-formation process. Numerous studies have characterized the relationship between star-formation activity, gas conditions, local environment, and galaxy morphology \citep[e.g.,][]{Kennicutt1998, Balogh2004, Kennicutt2012}. Volume-limited surveys \citep[e.g.,][]{Brinchmann2004, Lee2007, Bothwell2009} have mapped galaxy star-formation properties over stellar mass scales of $7\ltsimeq$ log (M$_*$/\msun) $\ltsimeq11$. These surveys have found that the properties vary monotonically over a luminosity range $-19 < $M$_B < -15$ mag, but the distribution of star forming properties widens at fainter magnitudes roughly corresponding to a stellar mass of log (M$_*$/\msun) of 8.5. Specifically, galaxies at the lower mass ranges of these larger surveys show a broadening in their star-formation and gas properties, with less correlation between gas mass, stellar mass, and star-formation activity. 

While these large surveys, as well as smaller, more detailed studies \citep[e.g., VLA-ANGST;][]{Ott2012}, have begun characterizing the star-formation and gas properties (and their dispersion) in lower mass galaxies, the faint end of the galaxy luminosity function remains largely unexplored. Outside of a dense group or cluster environment, these dwarf galaxies are expected to be gas rich, presumably evolving in relative quiescence due to inefficient star formation. Generally thought to have longer gas-cooling timescales and low, if not suppressed, star-formation activity, they play a key role in disentangling the factors driving star formation and star-formation driven galaxy evolution. Further highlighting the importance of low-mass galaxies, recent studies show that low-mass galaxies contribute significantly to both the total star-formation rate density at high redshift \citep[e.g.,][]{Alavi2014} and the ionizing background radiation \citep{Mostardi2013, Nestor2013}. While galaxies at the faint end of the luminosity function are predicted to be the most numerous class of galaxies, their intrinsic low luminosities and small angular sizes have made identifying and cataloging these systems through optical surveys difficult. Their high gas fractions make them better candidates for discovery through \HI\ surveys. 

The Arecibo Legacy Fast ALFA (ALFALFA) survey \citep{Giovanelli2005, Haynes2011} is a blind extragalactic \HI\ survey to map the nearby \HI\ universe over 7000 deg$^2$ of high Galactic latitude sky. With an \HI\ mass detection limit as low as $10^6$ \msun\ for galaxies in the Local Group and $10^{9.5}$ \msun\ at the survey velocity limit of $z\sim0.06$, the ALFALFA survey was designed to populate the faint end of the \HI\ mass function over a cosmologically significant volume.

The Survey of \HI\ in Extremely Low mass Dwarfs \citep[SHIELD;][]{Cannon2011} was designed to examine the early release of the ALFALFA dataset in a systematic investigation of nearby galaxies with HI masses $\ltsimeq 10^7$ \msun\ outside the Local Group. From the first $\sim10$\% of the processed ALFALFA survey data, 12 systems were selected for further study. $\textit{HST}$ optical imaging of these 12 systems have facilitated accurate distances based on the tip of the red giant branch (TRGB) method \citep{McQuinn2014}. The distances range from 5 to 12 Mpc; the \HI\ masses range from $4\times10^6-6\times10^7$ \msun\ based on these distances. All of the CMDs have red giant branch (RGB) sequences; a subset of the systems also show evidence of recent star formation based on their more populated upper main sequences (MS) and helium burning sequences \citep{Dohm-Palmer1997, McQuinn2011}. Most of the galaxies are detected in H$\alpha$ and spectroscopy of their \HII\ regions reveal that the SHIELD galaxies are metal poor \citep{Haurberg2014}.

In addition to providing accurate distance measurements, resolved stellar populations provide a means to measure the star-forming properties of a system by reconstructing a star formation history (SFH) using stellar evolution isochrones and color-magnitude diagram (CMD) fitting techniques \citep[e.g.,][]{Tosi1989, Tolstoy1996, Gallart1996, Dolphin1997, Holtzman1999, Harris2001}. This approach has been used to measure the star-formation rate as a function of time (SFR(t)) in a significant number of nearby galaxies, putting constraints on star formation, star-formation feedback, and galaxy evolution \citep[e.g.,][]{Gallart2007, Dalcanton2009, McQuinn2010a}. 

\begin{figure}
\epsscale{1.15}
\plotone{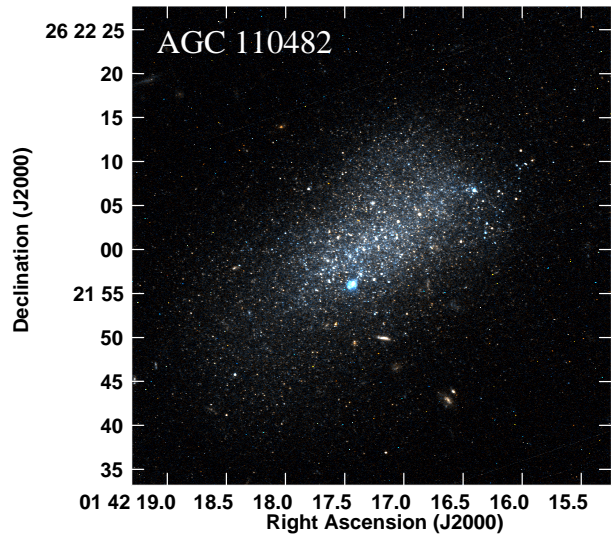}
\caption{$\textit{HST}$ composite image of AGC~110482, representative of the SHIELD sample, combining the F606W image (red), the average of F606W and F814W images (green), and the F814W image (blue). North is up and east is left.  The field of view encompasses twice the optical major axis as determined by iteratively plotting CMDs with different elliptical parameters.}
\label{fig:image}
\end{figure}

Here, we use the optically resolved imaging of the SHIELD galaxies to measure the star-formation characteristics of gas-rich galaxies in a mass range that is not well probed in previous studies. In this work, we use the photometry from $\textit{HST}$ optical images measured in \citet{McQuinn2014} to derive the star-formation properties using a CMD fitting technique and stellar evolution isochrones. In Section~2, we describe the sample, observations, and data reduction. In Section~3, we explain the CMD fitting technique used to measure the star-formation activity. In Sections~4 and 5, we present the lifetime average SFRs, recent SFRs, and discuss whether the star-formation activity appears to be correlated with the local environment. In Section~6, we compare the star-formation and stellar mass properties with the \HI\ gas properties previously measured from \citet{Cannon2011} and compare the results in these low-mass galaxies to previously published works. Finally, in Section~7 we summarize our conclusions. 

\section{The Galaxy Sample, Observations, and Photometry Method\label{data}}
Table~\ref{tab:galaxies} lists the 12 galaxies that comprise the SHIELD sample, their coordinates, observation details, and TRGB distances. The observations consist of optical imaging from the Advanced Camera for Surveys (ACS) Wide Field Channel \citep{Ford1998} aboard the $\textit{HST}$ in the F606W and F814W filters. New observations of 11 galaxies were obtained as part of the HST-GO-12658 SHIELD program (PI: Cannon). Archival observations were available for three galaxies from the previous HST-GO-10210 program (PI: Tully); two of these targets overlap with our new observations providing additional photometric depth to the final images. All observations were obtained with two filters during single $\textit{HST}$ orbits and were cosmic-ray split (CRSPLIT=2). Average integration times were 1000 s in the F606W filter and 1200 s in the F814W filter. The observations for the two galaxies that overlapped in the observing programs have longer final integration times of $\sim$1900 s in the F606W filter and $\sim$2300 s in the F814W filter. The observing programs and final integration times for each galaxy are noted in Table~\ref{tab:galaxies}.


\begin{deluxetable*}{lcccrrrrr}
\tablewidth{0pt}
\tablecaption{Summary of the SHIELD Sample and Observations\label{tab:galaxies}}
\tabletypesize{\scriptsize}
\tablecolumns{9}
\tablehead{
\colhead{}     			&
\colhead{}			&
\colhead{R.A.}         		&
\colhead{Decl.}		     	&
\colhead{HST}			&
\colhead{F606W}       		&
\colhead{F814W}         	&
\colhead{TRGB Dist}		&
\colhead{$A_{F606W}$}		
\\
\colhead{Galaxy}       		&
\colhead{Alt. Name}		&
\colhead{J2000}         	&
\colhead{J2000}         	&
\colhead{ID} 			&
\colhead{(sec)}       		&
\colhead{(sec)}	 		&
\colhead{(Mpc)}			&
\colhead{(mag)}			
\\
\colhead{(1)}      		&       
\colhead{(2)}        		&       
\colhead{(3)}      		&       
\colhead{(4)}        		&       
\colhead{(5)}        		&       
\colhead{(6)}          		&        
\colhead{(7)}          		&   
\colhead{(8)}          		&   
\colhead{(9)}			
}
\startdata
AGC~110482  & KK~13 & 01:42:17.4 & 26:22:00 & 12658 & 1008 & 1226 & 7.82$\pm0.21$	    & 0.26 \\
AGC~111164  & KK~17 & 02:00:10.1 & 28:49:52 & 10210 &  934 & 1226 & 5.11$\pm0.07$	    & 0.16 \\	   
AGC~111946  & KK~15 & 01:46:42.2 & 26:48:05 & 12658 & 1008 & 1126 & 9.02$_{-0.29}^{+0.20}$  & 0.24 \\     
AGC~111977  & KK~16 & 01:55:20.2 & 27:57:14 & 12658 & 1008 & 1226 & 5.96$_{-0.09}^{+0.11}$  & 0.20 \\	  
	    &       & 		 & 	    & 10210 &  934 & 1226 &			    &	   \\
AGC~112521  &       & 01:41:07.6 & 27:19:24 & 12658 & 1008 & 1226 & 6.58$\pm0.18$	    & 0.18 \\	  
AGC~174585  &       & 07:36:10.3 & 09:59:11 & 12658 & 1000 & 1220 & 7.89$_{-0.17}^{+0.21}$  & 0.11 \\	 
	    &       & 		 & 	    & 10210 &  900 &  900 &			    &	   \\
AGC~174605  &       & 07:50:21.7 & 07:47:40 & 12658 & 1000 & 1200 &10.89$\pm0.28$	    & 0.07 \\ 
AGC~182595  &       & 08:51:12.1 & 27:52:48 & 12658 & 1008 & 1226 & 9.02$\pm0.28$	    & 0.12 \\	  
AGC~731457  &       & 10:31:55.8 & 28:01:33 & 12658 & 1008 & 1226 &11.13$_{-0.16}^{+0.20}$  & 0.08 \\	   
AGC~748778  &       & 00:06:34.3 & 15:30:39 & 12658 & 1008 & 1226 & 6.46$_{-0.17}^{+0.14}$  & 0.19 \\	   
AGC~749237  &       & 12:26:23.4 & 27:44:44 & 12658 & 1008 & 1226 &11.62$_{-0.16}^{+0.20}$  & 0.06 \\	   
AGC~749241  &       & 12:40:01.7 & 26:19:19 & 12658 & 1008 & 1226 & 5.62$_{-0.14}^{+0.17}$  & 0.04 \\
\enddata
\tablecomments{\scriptsize{Column 1$-$Galaxy name. Column 2$-$Alternate galaxy name. Columns 3 and 4$-$Coordinates of galaxy in J2000. Column 5$-$HST observing program. Columns 6 and 7$-$Integration time in the F606W and F814W filters with the ACS instrument. Column 8$-$TRGB Distances from \citet{McQuinn2014}. The distances were derived using a Maximum-Likelihood technique; the uncertainties reflect the asymmetric distribution of photometric errors. Column 9$-$Galactic absorption from the dust maps of \citet{Schlegel1998} with the recalibration from \citet{Schlafly2011}.}}

\end{deluxetable*}

Figure~\ref{fig:image} shows an example composite image from the data of AGC~110482, reproduced from the previous study measuring the distances to the galaxies \citep[][see their Figure~1]{McQuinn2014}. The images were cosmic-ray cleaned, processed by the standard ACS pipeline, corrected for charge transfer efficiency non-linearities, and combined using \textsc{MultiDrizzle}. The F606W filter image is shown in blue, F814W in red, and the average of the F606W and F814W images in green. Typical of the sample, and of gas-rich dwarf galaxies in general, the image shows both a young, blue stellar population and a red, older and somewhat more extended stellar population. 

The sample has also been imaged in the narrowband H$\alpha$ filter from the WIYN 3.5 m telescope \citep{Haurberg2014}. These images reveal that 2 of the 12 galaxies lack \HII\ regions. Specifically, AGC~749241 was not detected in H$\alpha$ and AGC~748778 had only weak, diffuse H$\alpha$ emission but no \HII\ region. These H$\alpha$ data suggest that the star-formation activity over the most recent $\sim5$ Myr has been either below detection or very low (H$\alpha$-based SFR $< 10^{-5}$ \msun\ yr$^{-1}$). Thus, AGC~749241 and AGC~748778 meet the criteria of having both detectable amounts of \HI\ and no significant recent SFR as measured by H$\alpha$ and can be classified as transition dwarfs \citep[dTrans;][]{Mateo1998, Skillman2003, Weisz2011}. H$\alpha$-based SFRs for the rest of the sample will be presented in N.~C.~Haurberg (in preparation).

As described in detail in \citet{McQuinn2014}, photometry was performed on the pipeline processed, cosmic-ray cleaned images (CRJ files) using the ACS module of the DOLPHOT photometry package \citep{Dolphin2000}. Both quality and spatial cuts were performed on the photometry. Briefly, the photometry was filtered for well-fit point sources recovered with a signal-to-noise ratio $>5$ that were not severely affected by stellar crowding. Spatial cuts were determined iteratively by plotting the CMD of stars in concentric ellipses with the same ellipticity and position angles centered on the optical center of each galaxy. In the inner regions, the CMDs are dominated by stars from the galaxies, while in the outer regions, the CMDs are dominated by background sources. The semi-major and semi-minor axes of the ellipses were increased until the CMDs from larger annuli matched the distribution of point sources from a field region CMD. Artificial star tests using $\sim$500k stars were performed to measure the completeness limit of the images using the same photometry package and filtered on the same parameters. 

\begin{figure}
\epsscale{1.15}
\plotone{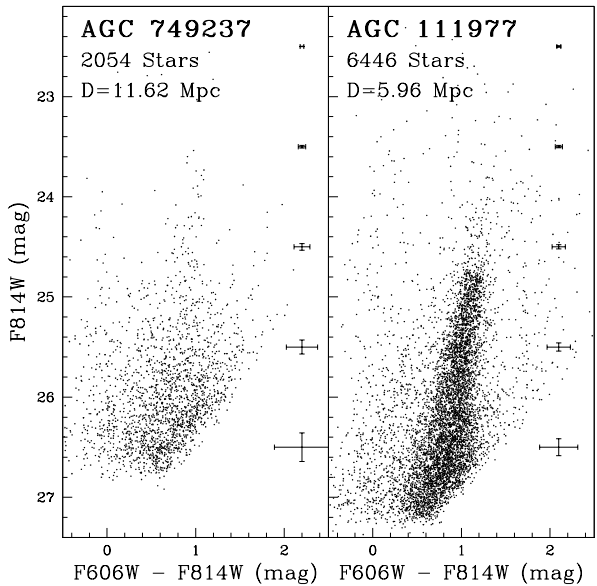}
\caption{CMDs of AGC~749237 (left) and AGC~111977 (right). Average uncertainties per magnitude are shown in each panel. The photometric depth of the CMD for AGC~749237 is typical of the 5 farthest galaxies, while the depth of the CMD for AGC~111977 is typical of the 7 closest galaxies. The number of stars in the CMDs across the sample varies between $\sim800$ and $\sim6500$.}
\label{fig:cmd}
\end{figure}

Figure~\ref{fig:cmd} presents an example CMD for one of the farthest galaxies from the SHIELD sample, AGC~749237, plotted to the 50\% completeness level as determined by the artificial star tests, and one of the closest galaxies, AGC~111977. Representative photometric uncertainties per magnitude are shown for both. The photometry in the CMDs were corrected for Galactic absorption based on the dust maps of \citet{Schlegel1998} with recalibration from \citet{Schlafly2011}; these values are also noted in Table~\ref{tab:galaxies}. The depth of photometry shown in Figure~\ref{fig:cmd} brackets that of the overall sample. For the 5 galaxies that are located between $8-12$ Mpc, the depth of photometry is similar to AGC~749237, reaching approximately 1 mag below the TRGB, but populated by differing numbers of stars. For the 7 galaxies that are located between $5-8$ Mpc, the depth of photometry is similar to AGC~111977, reaching approximately 2 mag below the TRGB. CMDs of all 12 galaxies can be found in \citet[][see their Figure 2]{McQuinn2014}. In each CMD, the MS is identifiable as well as a populated RGB sequence. The presence of stars in the upper MS indicates that a system has experienced star formation at recent times (t$<$ few 100 Myr), while the presence of the RGB stars indicates star-formation activity at older times (t$>1$ Gyr). Some of CMDs also show somewhat populated helium burning sequences, including AGC~110482, AGC~111977, AGC~731457, AGC~749237, which is a further indication of recent star-formation activity. Asymptotic giant branch (AGB) populations, indicative of intermediate age star-formation activity, are also observed in a subset of the CMDs, most notably in AGC~110482, AGC~174605, and AGC~182595. 

\section{Methodology for Deriving the Star-Formation Properties\label{sfhs}}
The star-formation characteristics of our sample of galaxies are reconstructed using stellar evolution isochrones to create a series of modeled CMDs from synthetic stellar populations of different ages and metallicities. The SFH of the best-fit modeled CMD to the observed CMD represents the most likely star-forming properties of the system. We use the numerical CMD fitting program, MATCH \citep{Dolphin2002}, with the stellar evolutionary models from \citet{Marigo2008} and updated AGB tracks from \citet{Girardi2010}. The primary inputs in reconstructing the past SFRs are the photometry and artificial star recovery fractions (i.e., quantitative observational uncertainties and incompleteness) coupled with the stellar evolutionary models. A Salpeter single sloped power law initial mass function with a spectral index of $-1.35$ from $0.1-120$~\msun\ \citep{Salpeter1955} is assumed as well as a binary fraction of $35\%$ with a flat secondary mass distribution. Distance is a free parameter fit by the SFH recovery program. We constrained the mean metallicity, $Z(t)$, to be a continuous, non-decreasing function with time. This physically motivated constraint guides the metallicity evolution in the absence of observational constraints that would be available with deeper photometric data and produces a more realistic metallicity evolution in the galaxies avoiding large jumps in metallicity over short time periods.

Photometric errors and extinction can broaden features in a CMD and are explicitly accounted for in the CMD fitting program. The photometric uncertainties and completeness are quantified using the artificial star tests. Extinction is a free parameter fit by the SFH recovery program. Thus, while the photometry shown in the CMDs in Figure~\ref{fig:cmd} has been corrected for foreground extinction, we used the uncorrected photometry in the CMD fitting program. Both foreground and internal extinction are expected to be low for this sample. The galaxies are located at high Galactic latitudes with foreground extinction estimated to range from $A_{F606W}$ of 0.04 to 0.26 mag (see Table~\ref{tab:galaxies}). Internal extinction is also expected to be low as dwarf galaxies have been shown to follow the mass-metallicity relation \citep{Berg2012}. Nonetheless, similar to the distance, extinction is an important parameter to fit in measuring the SFHs and provides a consistency check when compared to independently measured values. For our sample the distances and extinctions derived from the best-fit CMDs were within 0.1 mag and 0.05 mag, respectively, of the measured TRGB distances and foreground extinction estimates. Once the best fit has been verified, the final SFH solution is derived with the distance and extinction parameters fixed to the measured values from Table~\ref{tab:galaxies}.

\begin{figure*}
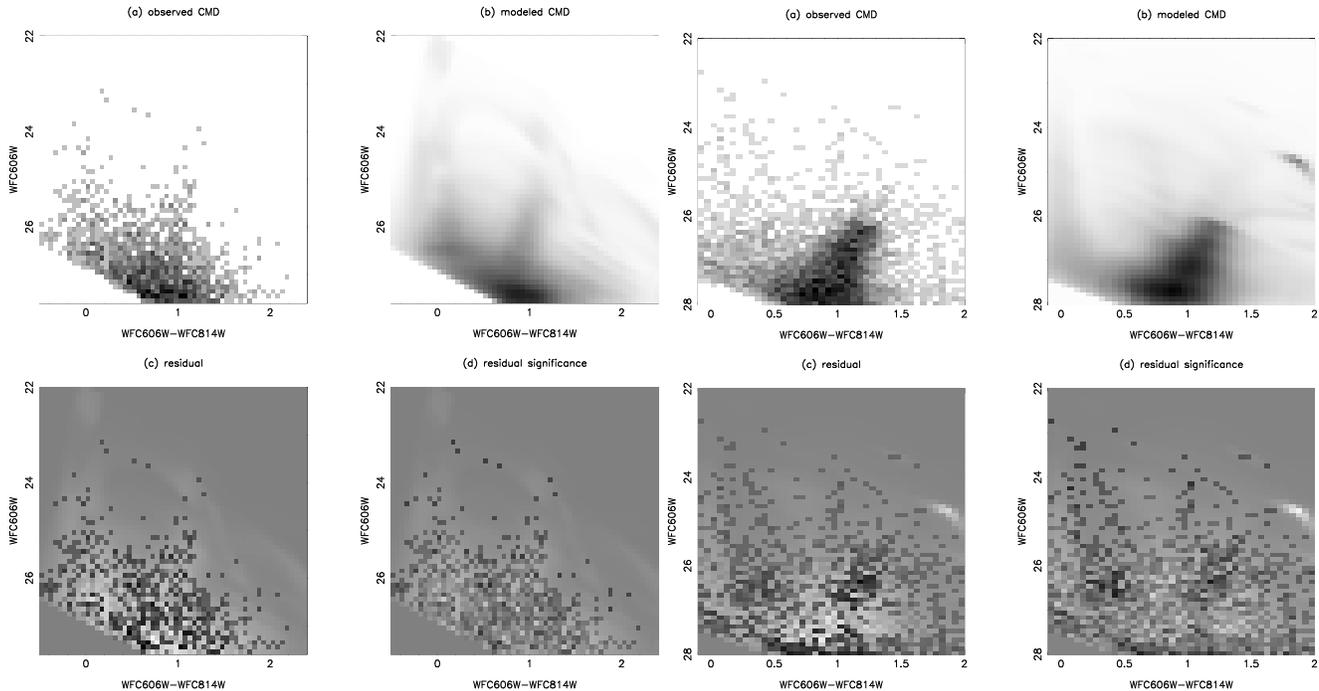

\includegraphics[width=0.48\linewidth]{f3a.eps}
\includegraphics[width=0.48\linewidth]{f3b.eps}
\caption{$\textit{Left 4 panels:}$ AGC~749237: Hess diagrams of the observed CMD compared with the best-fit modeled CMD, residual of the data$-$model CMD with black and white points representing $\pm5\sigma$, and the residual significance CMD which weights the data$-$model by the variance in each Hess bin \citep{Dolphin2002}. The synthetic CMD reconstructs the different features seen in the observed CMD with minor discrepancies. The axes labels combine the name of the HST instrument (ACS WFC) with filter number. $\textit{Right 4 panels:}$ AGC~111977: Same sequence of hess diagrams as shown for AGC~749237.}
\label{fig:hess}
\end{figure*}

Figure~\ref{fig:hess} presents two examples of the observed CMDs alongside the modeled CMD on the same axis scale for comparison as well as the residuals and weighted residuals between the observed and modeled CMD. As seen in Figure~\ref{fig:hess}, the models characterize all of the features of the CMDs quite well with only small discrepancies. The synthetic CMD of the galaxy is based on the most likely SFH and metallicity of the galaxy given our inputs and models \citep[e.g.,][]{Dolphin2003}. While the constraints on the metallicity evolution of the sample are limited by the photometric depth of the data, the best-fit values of the most-recent metallicities are in agreement with spectroscopic oxygen abundance measurements and the mass-metallicity relation for low-mass galaxies as discussed in \citet{Haurberg2014}, providing an additional consistency check on our results. 

Uncertainties on the SFHs take into account systematic and random uncertainties. The systematic uncertainties measure the uncertainty in the SFHs due to variations of the real data from the theoretical stellar evolution models \citep{Dolphin2012}. These uncertainties are estimated by applying shifts to the models in both bolometric luminosity and temperature using Monte Carlo simulations. Random uncertainties were estimated using a new method of applying a hybrid Markov Chain Monte Carlo simulation as described by \citet{Dolphin2013}. This technique samples the probability distribution of the parameters used in the original solution to directly estimate confidence intervals. The inclusion of this additional source of uncertainty is particularly important for time bins where the SFR is low or zero, and for data whose photometric depth is above the horizontal branch or red clump as is the case for the current galaxy sample. The final uncertainties in the SFHs are calculated by adding the uncertainties in quadrature.

Reconstructing detailed SFHs (i.e., SFRs(t)) requires sufficient information on the ages and metallicity of the stellar populations from a CMD. Therefore, the photometric depth of the data is the main determinant on the temporal resolution achievable in a SFH at older ages. For a discussion on the systematic effects of photometric depth on measuring a SFH see \citet{McQuinn2010a}. A secondary determinant on the temporal resolution is the number of stars populating a CMD, which becomes a more significant factor to consider in low-mass galaxies. The small number of stars in the SHIELD galaxies compared to more massive dwarfs \citep{Dalcanton2009, McQuinn2010a} is therefore one of the dominant limitations in determining recent SFHs with high time resolution. Through extensive tests of MATCH using different time binning schemes, we found that measuring the SFRs over two timescales, recent (t$\sim$200 Myr; primarily measured by the upper part of an optical CMD) and lifetime-average SFRs (primarily measured from the RGB stars), resulted in the most robustly measured SFRs across the sample. The recent timescale of 200 Myr is an appropriate choice for the entire sample, regardless of distance and photometric depth, because of the generally low star-formation activity in low-mass galaxies. Further, the $\sim$200 Myr timescale for the recent SFRs is comparable to the timescale measured by near UV studies of star-formation rates, in particular from GALEX images \citep[e.g.,][]{Hao2011}. Thus, the recent SFR measurements reported here will enable comparisons with a large number of galaxies that have SFRs measured from integrated UV light. 


\begin{deluxetable*}{lcccllcl}
\tablewidth{0pt}
\tablecaption{Star Formation and Gas Properties\label{tab:sf}}
\tabletypesize{\scriptsize}
\tablecolumns{8}
\scriptsize
\tablehead{
\colhead{}                      	&
\colhead{SFR}    	 	&
\colhead{SFR}    	 	&
\colhead{Recent/}              &
\colhead{Total}			&
\colhead{}				&
\colhead{Gas}			&
\colhead{}
\\
\colhead{}				&
\colhead{$<200 Myr>$}	&
\colhead{$<Lifetime>$}      &
\colhead{Lifetime}		&
\colhead{M$_*$}                &
\colhead{M$_{HI}$}		&
\colhead{Ratio}			&
\colhead{$\tau_{gas}$}			
\\
\colhead{Galaxy}                &
\colhead{($10^{-3}$ \msun\ yr$^{-1}$)}&
\colhead{($10^{-3}$ \msun\ yr$^{-1}$)}&
\colhead{SFR}    	 	&
\colhead{(10$^6$ \msun)} 	&
\colhead{(10$^6$ \msun)}	&
\colhead{(M$_{HI}$/M$_*$)}&
\colhead{(Gyr)}		
\\
\colhead{(1)}			&
\colhead{(2)}			&
\colhead{(3)}			&
\colhead{(4)}			&
\colhead{(5)}			&
\colhead{(6)}			&
\colhead{(7)}			&			
\colhead{(8)}			
}
\startdata
AGC~110482  &  5.4 $^{+2.3}_{-2.2}$   	&  5.5 $^{+1.9}_{-2.0}$   	& 1.0 $\pm0.5$			& 55  $^{+19}_{-19}$ & 19.$\pm2$	& 0.35$\pm0.13$ 		& 6.7$\pm2.9$  	\\
AGC~111164  &  0.52 $^{+0.23}_{-0.35}$	&  1.0 $^{+0.2}_{-0.3}$   	& 0.5 $^{+0.3}_{-0.4}$	& 10  $^{+2}_{-3}$   	& 4.0$\pm0.4$	& 0.42$^{+0.09}_{-0.15}$ 	& 14$^{+7}_{-10}$	\\
AGC~111946  &  3.5 $^{+1.7}_{-2.1}$   	&  1.7 $^{+0.6}_{-0.7}$   	& 2.1 $^{+1.3}_{-1.5}$ 	& 17  $^{+6}_{-7}$   	& 15$\pm1.7$  	& 0.88$^{+0.35}_{-0.38}$ 	& 7.8$^{+4.0}_{-4.7}$\\
AGC~111977  &  2.6 $^{+1.3}_{-1.1}$   	&  3.8 $^{+1.2}_{-1.1}$   	& 0.7	 $^{+0.4}_{-0.4}$	& 37  $^{+12}_{-11}$& 7.1$\pm0.8$ 	& 0.19$\pm0.06$ 		& 5.3$^{+2.7}_{-2.4}$\\
AGC~112521  &  $\leq0.36^{+0.14}$    	&  0.75 $^{+0.26}_{-0.25}$& $\leq0.5^{+0.3}$		& 7    $^{+ 3}_{-2}$ 	& 7.0$\pm0.8$ 	& 0.94$^{+0.35}_{-0.33}$ 	& $\geq37^{+15}$  	\\
AGC~174585  &  2.3 $^{+1.2}_{-1.3}$   	&  0.90 $^{+0.33}_{-0.30}$& 2.5 $^{+1.6}_{-1.6}$	& 9   $^{+ 3}_{- 3}$   & 7.9$\pm0.9$  & 0.89$^{+0.34}_{-0.32}$  & 6.6$^{+3.5}_{-3.7}$\\
AGC~174605  &  4.3 $^{+1.7}_{-2.5}$   	&  $\leq2.8^{+1.4}$   	 	& $\geq1.5^{+1.0}$		& $\leq28^{+14}$ 	& 19$\pm2$ 	& $\geq0.66^{+0.33}$ 	& 8.2$^{+3.4}_{-4.9}$\\
AGC~182595  &  4.8 $^{+3.0}_{-1.2}$   	&  5.1 $^{+2.3}_{-3.3}$   	& 0.9 $^{+0.7}_{-0.6}$	& 50  $^{+22}_{-32}$ & 8.1$\pm0.9$ 	& 0.16$^{+0.07}_{-0.10}$ 	& 3.2$^{+2.1}_{-0.9}$\\
AGC~731457  &  14. $\pm6.$      	    	&  6.6 $^{+3.7}_{-4.9}$   	& 2.1 $^{+1.5}_{-1.8}$	& 65  $^{+37}_{-48}$ & 18$\pm2$ 	& 0.28$^{+0.16}_{-0.21}$ 	& 2.5$\pm1.1$ 		\\
AGC~748778  & 0.68 $^{+0.26}_{-0.38}$	&  0.32 $^{+0.06}_{-0.11}$	& 2.1 $^{+0.9}_{-1.4}$ 	&   3  $^{+ 1}_{- 1}$ 	& 4.5$\pm0.5$ 	& 1.4$^{+0.32}_{-0.54}$ 	& 13$^{+5}_{-7}$  	\\
AGC~749237  &  13. $^{+7.}_{-2.}$       	&  $\leq5.4^{+2.9}$   		& $\geq2.4^{+1.8}$		& $\leq53^{+29}$ 	& 57$\pm6$ 	& $\geq1.1^{+0.60}$ 		& 8.4$^{+4.5}_{-1.6}$\\
AGC~749241  & $\leq0.32^{+0.08}$      	&  0.36 $^{+0.13}_{-0.24}$& $\leq0.9^{+0.4}$		&   4 $^{+ 1}_{- 2}$ 	& 5.7$\pm0.7$ 	& 1.6$^{+0.60}_{-1.1}$ 	& $\geq34^{+9}$ 	\\ 
\\
\enddata

\tablecomments{\scriptsize{Column 1$-$Galaxy name. Columns 2$-$3 Average SFRs over the past 200 Myr and the lifetime of the galaxy derived from the CMD fitting technique. Uncertainties include both systematic and random uncertainties; see text for details. For cases where only an upper limit on the SFR could be determined, no lower bound on the uncertainty is listed. Column 4$-$Ratio of the recent to lifetime average SFR. Columns 5$-$ Stellar Masses formed over the lifetime of the galaxies, calculated from the SFRs in column 3 and assuming 30\% of the stellar mass is recycled over the lifetime of the galaxy \citep{Kennicutt1994}. Column 6$-$\HI\ mass based on the \HI\ flux measurements from \citet{Cannon2011} and the adopted TRGB distances from \citet{McQuinn2014}. Note the \HI\ masses reported for AGC~112521 and AGC~182595 have been updated. Columns 7$-$ Gas Ratio (M$_{HI}$ / M$_*$). Column 8$-$Gas consumption timescales based on the recent SFRs, assuming a 30\% recycling fraction of mass, and the \HI\ masses with a 1.33 correction for helium. }}
\end{deluxetable*}

Note that the data for a few of the galaxies allow only upper limits to be placed on the SFRs. For the lifetime SFRs, this is mainly due to limits in photometric depth and impacts two of the three farthest galaxies. For the more recent SFRs, this is a result of the low number of young stars available to fit in the CMD and impacts the two galaxies with the lowest recent SFRs. The recent and lifetime SFRs, as well as the uncertainties, are reported in Table~\ref{tab:sf}. For the galaxies whose measured SFRs are upper limits, we report only the upper bound of the uncertainties.

Stellar mass estimates can be derived from the lifetime-average SFRs. Integrating the lifetime-average SFR over the life of a galaxy yields the total stellar mass created in a system. This total stellar mass can be converted to a measurement of the current stellar mass by applying a correction for the amount of mass returned to a galaxy over the lifetime of the stellar populations. We used the recycling model from \citet{Kennicutt1994} who estimate that 30\% of the mass in a stellar generation is returned to a galaxy over its lifetime based on a Salpeter IMF. These values are reported in Table~\ref{tab:sf}. As a consistency check on our stellar mass values, we compared our results with preliminary estimates of the stellar masses based on mass-to-light (M/L) ratio analysis using \textit{Spitzer} Space Telescope IRAC imaging (J.~M.~Cannon et al., in preparation). Our stellar mass estimates are a factor of 2 higher than the majority of the stellar mass estimated from the M/L ratios, but are still in general agreement with one another given the large uncertainties. The exception is AGC~748778 whose stellar mass estimate from the M/L ratio is significantly lower than our value.  Future work on the SHIELD sample will include a larger discussion of the two stellar mass estimates including a description of the infrared imaging, analysis, and uncertainties (J.~M.~Cannon et al., in preparation).

\begin{figure}
\epsscale{1.15}
\plotone{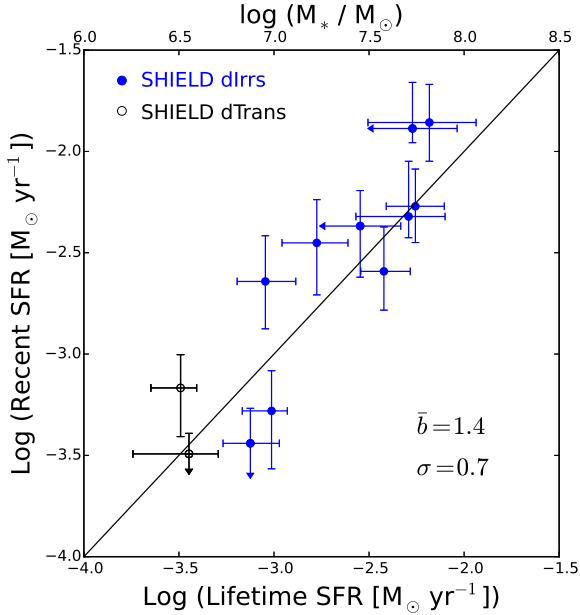}
\caption{Recent (t$<200$ Myr) SFRs compared with average lifetime SFRs for the SHIELD sample. To provide further context, the scale on the top x-axis is stellar mass based on the lifetime average SFRs and the age of the Universe and assuming a 30\% recycling fraction. Arrows denote cases where upper limits were placed on the SFRs. The solid line represents equivalent recent and lifetime SFRs. Overall, the recent SFRs are comparable to, or slightly higher than, the lifetime average SFRs. The birthrate parameters fall in a narrow range with a mean of 1.4 and a dispersion of 0.7. The level of recent star-formation activity in each system does not appear to be correlated with distance from the nearest known galactic neighbors or local environments.}
\label{fig:sfrs_stellar_mass}
\end{figure}

\begin{figure}
\epsscale{1.15}
\plotone{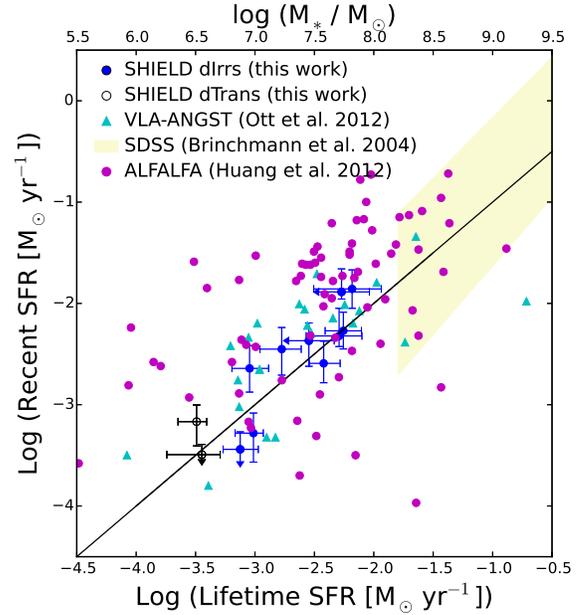}
\caption{An expanded view of Figure~\ref{fig:sfrs_stellar_mass} with results from 3 different surveys over-plotted. From \citet{Ott2012}, typical uncertainties for the VLA-ANGST recent SFRs are of order $\pm0.2$ dex; typical uncertainties for stellar mass measurements were not provided. From \citet{Huang2012}, typical uncertainties are of order $\pm0.3$ dex for the recent SFRs, and $\pm0.2$ dex for the stellar mass measurements. The shaded region for the SDSS survey covers the general range of properties that overlap with the other surveys plotted. The recent and lifetime star-formation measurements are consistent with those measured by these three surveys in the overlapping mass range. Further, the dTrans are indistinguishable from the dIrr galaxies. These low-mass galaxies follow the general trends that lower-mass galaxies have lower SFRs.}
\label{fig:sfrs_angst}
\end{figure}

\section{Star-Formation Rates and Comparison with Other Surveys\label{sfrs}}
Figure~\ref{fig:sfrs_stellar_mass} presents a comparison between the recent (t$<200$ Myr) SFRs with the average lifetime SFRs for the SHIELD sample. dIrrs are plotted as filled, blue cirlces; dTrans are plotted as unfilled, black cirlces. The recent and lifetime SFRs are low, ranging from $-3.5  \ltsimeq$ log (SFR) $\ltsimeq -1.8$, with SFRs in units of \msun\ yr$^{-1}$. The top axis in Figure~\ref{fig:sfrs_stellar_mass} shows the stellar mass for each system, which provides a complementary view of the results. Seen in this way, it is clear that the galaxies follow the well-established correlation that higher mass galaxies have higher recent SFRs. This trend reflects that while gas fractions tend to decline with increasing galaxy mass, total gas masses increase as a function of mass and can thus support higher SFRs \citep[e.g.,][]{Boselli2001}.

In Figure~\ref{fig:sfrs_stellar_mass}, the solid line marks where the recent to lifetime SFR ratio \citep[i.e., the birthrate parameter $b \equiv SFR_{recent}/SFR_{lifetime}$;][]{Scalo1986} is equal to 1. The birthrate parameters measured for the sample are listed in Table~\ref{tab:sf}. The range in the b parameter is narrow with a mean value of 1.4 and a dispersion of 0.7. To test whether the dispersion is due to measurement errors or to intrinsic scatter in star-formation activity, we fit a linear relationship to the recent and lifetime SFRs using the IDL mpfitexy routine \citep{Williams2010}. The mpfitexy routine depends on the mpfit package \citep{Markwardt2009}. We find that less than 35\% of the dispersion is due to intrinsic scatter while greater than 65\% is due to the measured uncertainties. Overall the SHIELD sample shows recent star-formation activity that is comparable to, or slightly higher than, the lifetime average SFR. Further, none of the galaxies show star-formation activity that is significantly lower than their lifetime average (i.e., no quenched star formation), as might be expected from their relatively low-density environments \citep{Hodge1971, Einasto1974, vandenBergh1994a, vandenBergh1994b, Mateo1998, McConnachie2012, Geha2012}. 

Figure~\ref{fig:sfrs_angst} expands the plot from Figure~\ref{fig:sfrs_stellar_mass} to include results from three other surveys. We add the recent SFRs and stellar masses from both the VLA-ANGST sample \citep{Ott2012} and the sub-sample of gas-rich ALFALFA dwarf galaxies from \citet{Huang2012}, and highlight the general range of properties measured in the SDSS survey from \citet{Brinchmann2004} that overlaps with the parameter space in Figure~\ref{fig:sfrs_angst}. All stellar masses were normalized to a Salpeter IMF to facilitate the comparison. Note however, that the stellar masses from these surveys were derived using very different techniques including mass-to-light ratios \citep{Ott2012}, stellar absorption-line indices \citep{Brinchmann2004}, and spectral energy distribution (SED) fitting \citep{Huang2012}. In addition to the $\sim50$\% uncertainties in stellar mass from these various techniques, there are also systematic uncertainties between techniques that are not considered. However, these uncertainties are less than the broad parameter space probed in Figure~\ref{fig:sfrs_angst}, making the comparison between the surveys of value. The recent SFRs have also been measured using different techniques. The SFRs from \citet{Brinchmann2004} were based on emission line model fits and integrated H$\alpha$ emission measurements scaled to a SFR. The SFRs from \citet{Ott2012} were derived by scaling integrated ultraviolet luminosities. These UV-based SFRs benefit from tracing the star-formation activity on comparable timescale to our CMD-based recent SFRs. The SFRs in \citet{Huang2012} were derived from SED fitting and were shown to agree reasonably well with SFRs derived by scaling integrated ultraviolet luminosities in systems with low extinction. 

\begin{figure}
\epsscale{1.15}
\plotone{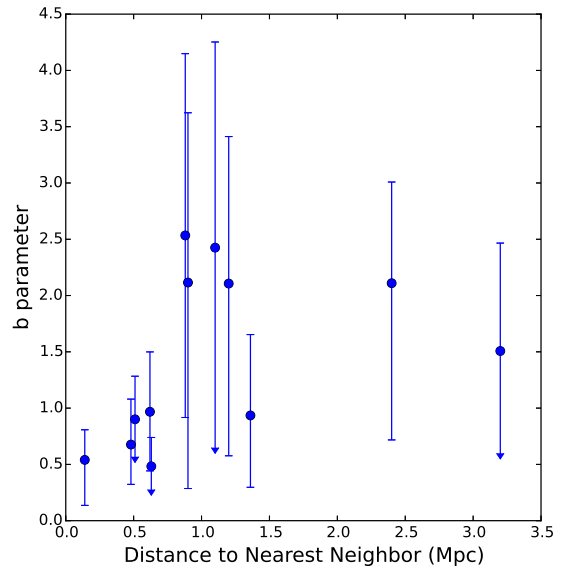}
\caption{The birthrate parameter is plotted against the distance to the nearest known neighbor from \citet{McQuinn2014}. While the overall range in the b parameter is rather narrow, the galaxies with the closest known neighboring system have the lowest values of the birthrate parameter, while the galaxies that are located at larger distances from their nearest neighbor have higher values of the birthrate parameter, albeit with large uncertainties.}
\label{fig:neighbor_b}
\end{figure}

There are a number of correlations to note in Figure~\ref{fig:sfrs_angst}. First, while the dTrans (unfilled black points) are at the low-mass end of the SHIELD sample, their star-formation properties are consistent with the rest of the SHIELD sample and are indistinguishable from the properties of the dIrrs galaxies measured from the other surveys. Second, the trend that gas-rich low-mass galaxies generally have higher recent SFRs compared with lifetime average SFRs (i.e., b $\gtsimeq$ 1) is more apparent across the combined samples. Third, while the properties of the SHIELD galaxies are consistent with these other studies, the SHIELD galaxies have a narrower range in birthrate parameters, particularly compared to the ALFALFA dwarfs from \citet{Huang2012}. One possible explanation is that the broader range in the birthrate parameter in \citet{Huang2012} is due to higher uncertainties in comparing the stellar masses and SFRs by the aforementioned different techniques. On the other hand, the SHIELD sample is small; future work will expand the SHIELD sample to $\sim$30 galaxies and may help determine whether the narrow range in recent-to-lifetime SFRs is typical in this very low-mass regime (K.~B.~W.~McQuinn et al. in preparation). Note that the opposite trend (i.e., lower recent to historical average SFRs) was found in \citet{Skillman2003} for a sample of gas-rich dwarf galaxies in the Sculptor group. However, the SFRs were based on integrated H$\alpha$ emission measurements which may not accurately represent the recent star-formation activity. SFRs based on H$\alpha$ emission trace only the most recent ($\sim$5 Myr) star-formation activity and depend both on constant levels of star formation and on a fully-populated upper-end of the IMF. In this low SFR regime, not only can the star formation fluctuate on short timescales, but the IMF can be stochastically populated \citep{Boselli2009, Lee2009, Goddard2010, Koda2012}. Thus, the H$\alpha$-based SFRs can be an unreliable tracer of star-formation activity, introducing scatter in the measurements and potentially biasing SFR measurements low in dwarf galaxies.

\section{Recent SFRs: Influenced by Environment or Regulated Internally?\label{recent_sfrs}}
One of the outstanding questions concerning the evolution of relatively isolated, low-mass galaxies is the comparative importance of environmentally driven star-formation activity versus star-formation activity regulated by internal conditions within a system. The former is directly correlated with the number of, and proximity to, neighboring systems. A full investigation of the role environment plays requires a complete census of a galaxy's environment including positional and velocity information of neighboring systems and SFHs with fine temporal resolution. However, to first order, by comparing the recent SFRs of galaxies derived here with the proximity of their known neighboring systems determined from our previous study \citep[][see their Figures~5 and 6]{McQuinn2014}, we can probe whether there is an obvious environmental factor influencing their recent evolution. 

In Figure~\ref{fig:neighbor_b} we present a comparison of the b values for the sample with the distances to the nearest known neighbors from \citet{McQuinn2014}. All six galaxies whose recent star-formation activity is somewhat higher than their lifetime average lie in low-density environments. Specifically, AGC~748778, AGC~174605, and  AGC~749237 are truly isolated with no known neighbors identified within a 1 Mpc radius. AGC~174585 and AGC~731457 are both located $\sim0.9$ Mpc from their nearest known neighbors, namely the $14+19$ association and the low-mass galaxy DDO~83 respectively. AGC~111946 is located at the end of a linear structure comprised of two known galaxy associations, the NGC~784 and NGC~672 groups \citep{Zitrin2008, McQuinn2014}. While AGC~111946 appears to be part of this elongated 4 Mpc structure, it is separated by a distance of $\sim1$ Mpc from the more tightly clustered galaxies in the NGC~672 group. Overall, the isolated nature of these six galaxies suggests that the recent star-formation activity is not driven by gravitational interactions. Note however that we cannot rule out the possibility of possible undetected gas-poor companions outside of the HST fields of view.

Further, the three galaxies that show somewhat lower levels of recent star-formation activity compared with their lifetime averages (AGC~111164, AGC~111977, and AGC~112521) are part of the galaxy associations of the NGC~784 and NGC~672 groups mentioned above. This linear structure of galaxies stretches $\sim4$ Mpc and is thought to trace the local matter distribution. In contrast to the location of the aforementioned system AGC~111946, these three SHIELD systems lie at closer distances to the members of these groups. The nearest neighbor to AGC~111164 is the more massive dwarf starburst galaxy NGC~784 at a distance of $\sim0.14$ Mpc. The SFR in NGC~784 has been noted to be elevated over the last few 100 Myr \citep{McQuinn2010b}. However, despite being separated by only $\sim0.14$ Mpc, the recent SFR in AGC~111164 is half its lifetime average. AGC~111977 is located $\sim0.65$ Mpc from AGC~112521, which is also separated by a comparable distance in the opposite direction from four members of the NGC~672 group, namely NGC~672, IC~1727, AGC~111945, and the SHIELD galaxy AGC~110482. Even though these galaxies reside in a more populated environment, their recent star-formation activity is lower than their lifetime values. Note that the \HI\ masses for these galaxies are on the low end of our sample ($4\times10^6$ \msun\ for AGC~111164 and $7\times10^6$ \msun\ for AGC~111977 and AGC~112521), which may provide a partial explanation for the low recent star-formation activity. 

Finally, we examine the three galaxies whose recent star-formation activity is approximately equivalent to their lifetime averages, namely AGC~110482, AGC~182595, and AGC~749241. The first system, AGC~1110482, is located in the galaxy association described above, $\gtsimeq0.65$ Mpc from other members of the NGC~672 group. AGC~182595 is truly isolated with no known neighbors with 1 Mpc. AGC~749241 lies in a more populated region within 9 other galaxies which form a linear structure that extends 1.6 Mpc from end to end. Given the distribution of galaxies in this region, it is likely that AGC~794241 has been influenced by the external gravitational perturbations from other systems. This is supported by the irregular \HI\ morphology whose centroid is not co-spatial with ground-based optical imaging of the stellar population \citep[][see their Figure 2]{Cannon2011}. However, the relatively low-level of recent star-formation activity in AGC~749241 indicates that any such gravitational interaction has not had a dramatic impact on the star forming properties over the past 200 Myr. 

\begin{figure}
\epsscale{1.15}
\plotone{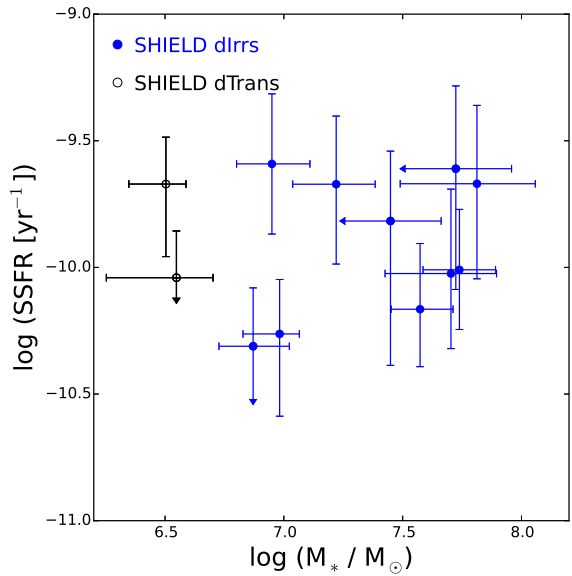}
\caption{Specific SFRs as a function of stellar mass. In contrast to the star-forming properties in more massive galaxies, there is a high degree of dispersion in these very low-mass galaxies. This dispersion likely reflects small fluctuations in star-formation activity superposed on already low average SFRs.}
\label{fig:ssfr_stellar_mass}
\end{figure}

\begin{figure}
\epsscale{1.15}
\plotone{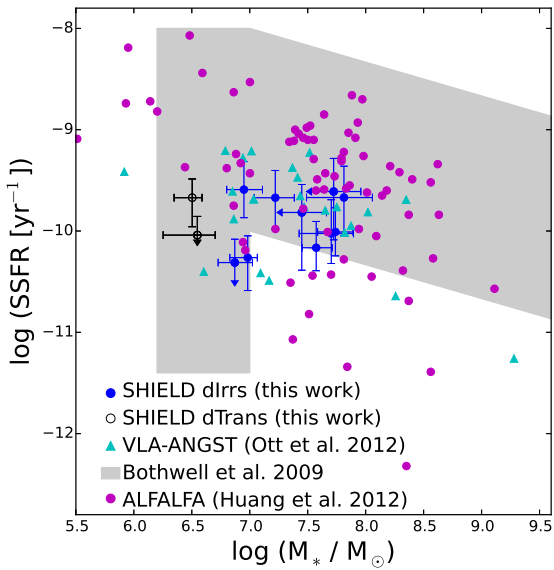}
\caption{An expanded view of Figure~\ref{fig:ssfr_stellar_mass} with results from 3 different surveys over-plotted. Uncertainties for the SSFR measurements are greater than $\pm0.2$ dex for VLA-ANGST and $\pm0.3$ dex for ALFALFA; uncertainties for the stellar mass were not provided for the VLA-ANGST measurements and are of order $\pm0.2$ dex for ALFALFA. The shaded region from \citet{Bothwell2009} covers the range of points included in their study; the extension to lower SSFRs includes both measurements and upper limits. Overall, for low-mass galaxies none of the samples show a high degree of correlation between the SSFRs and stellar mass.}
\label{fig:ssfr_angst}
\end{figure}

In summary, while the range in b parameters is fairly small, the somewhat higher levels of recent star-formation activity in these low-density galaxy environments do not correlate with smaller distances to the nearest neighboring galaxies, or vice versa, suggesting that the recent activity is governed by internal conditions and not environmentally driven. Similarly, \citet{Hunter2004} did not find a correlation between star-formation activity and distances to nearest neighbors in a larger sample of gas-rich dwarf galaxies. However, these authors note the possibility that this could be a selection effect as there could be undetected low-mass neighbors or \HI\ clouds near the galaxies with higher measured recent SFRs. While we cannot rule out the possibility of undetected neighboring systems, particularly if they are gas-poor, there are no  systems detected nearby in position or velocity in the ALFALFA catalog \citep{Haynes2011}. Further, no nearby systems have been found in the VLA data of the sample \citep{Cannon2011}, nor in the HST images which cover a field of view extending to $\sim0.5$ Mpc for the farther systems \citep{McQuinn2014}. 

For the galaxies with higher recent SFRs, an alternative explanation to star formation driven by gravitational perturbations may be that these low-mass galaxies have experienced a delayed onset to star formation, similar to 3 low-mass galaxies found in the Local Group (Leo~A, Leo~T, and Aquarius). In a detailed study of the ancient star formation history of Leo~A, \citet{Cole2007} found that over 90\% of the stars in the galaxy were formed over the most recent 8 Gyr. In a similar study of Leo~T, \citet{Weisz2012} found $\sim50$\% of the stars in the galaxy were formed over the same timescale. The Aquarius dIrr also shows delayed onset of star-formation with 90\% of the stars forming in the last 10 Gyr \citep{Cole2014}. While a limited comparison based only on three galaxies, these examples of delayed or suppressed star formation in low-mass galaxies lend support to the idea of galaxy downsizing. Higher recent star-formation activity in low-mass galaxies may not always be tied to gravitational interactions, but may instead be the result of a combination of internal properties and conditions governed by, for example, gas cooling timescales or reionization, and stochastic phenomena such as mergers.

\section{Comparison of Star-Formation with Global Gas Properties\label{properties}}
The recent star-formation activity of the sample can also be analyzed by normalizing the SFR by the stellar mass (M$_*$) of each galaxy. Figure~\ref{fig:ssfr_stellar_mass} shows the specific star-formation rates (SSFR $\equiv$ recent SFR / M$_*$) as a function of stellar mass. There is no apparent correlation between the SSFR and the stellar mass of the galaxies. This is in agreement with previous surveys of low-mass galaxies which have noted a high degree of scatter in the SSFRs as a function of stellar mass in systems with SFRs $<10^{-3}$ \msun~yr$^{-1}$ \citep[e.g.,][]{Bothwell2009, Huang2012}. 

Figure~\ref{fig:ssfr_angst} expands on Figure~\ref{fig:ssfr_stellar_mass} by adding results from the VLA-ANGST survey \citep{Ott2012} and a different sub-sample of gas-rich dwarf galaxies from the ALFALFA survey \citep{Huang2012}, as well as adding the range of galaxy properties measured in the survey by \citet{Bothwell2009}. The scatter is the greatest in the study by \citet{Bothwell2009}, particularly below M$_* < 10^7$ \msun, which includes a number of upper limits on galaxies with no detectable star formation. Similarly to the SFRs from \citet{Skillman2003}, the SFRs in \citet{Bothwell2009} were based on integrated H$\alpha$ emission which can be unreliable in the low-SFR regime. As a result, \citet{Bothwell2009} was unable to determine whether the scatter in the SSFR-stellar mass relation was intrinsic, or due to the H$\alpha$ emission not adequately tracing the SFR. \citet{Huang2012} also find a large scatter in SSFRs for low-luminosity galaxies, including a similar quenching of star formation for some galaxies with $M_* < 10^8$ \msun. However, since the SFRs in this study were based on SED fitting, \citet{Huang2012} conclude that the range in SSFR, including the departure to low SSFRs for some low-mass galaxies, is real.  In comparison, the SHIELD and VLA-ANGST samples show a narrower scatter in the SSFR-stellar mass relationship, and none of the galaxies in the SHIELD sample show quenched star formation (i.e., none have outlying low values of SSFR). 

\begin{figure}
\epsscale{1.15}
\plotone{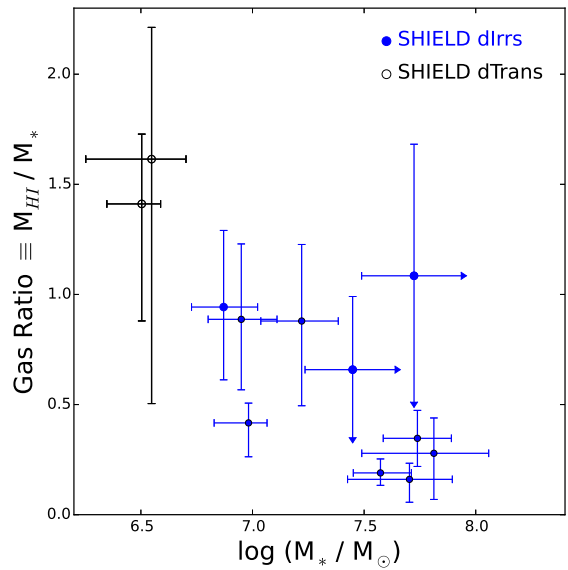}
\caption{Fraction of \HI\ to stellar mass vs. stellar mass for the 12 SHIELD galaxies. The stellar mass was determined from the lifetime average SFRs of the galaxies as described in the text. The \HI\ mass was calculated from the \HI\ fluxes reported in \citet{Cannon2011}. The range in gas fractions is consistent with more massive gas-rich dIrrs.}
\label{fig:gas_frac}
\end{figure}

\begin{figure}
\epsscale{1.15}
\plotone{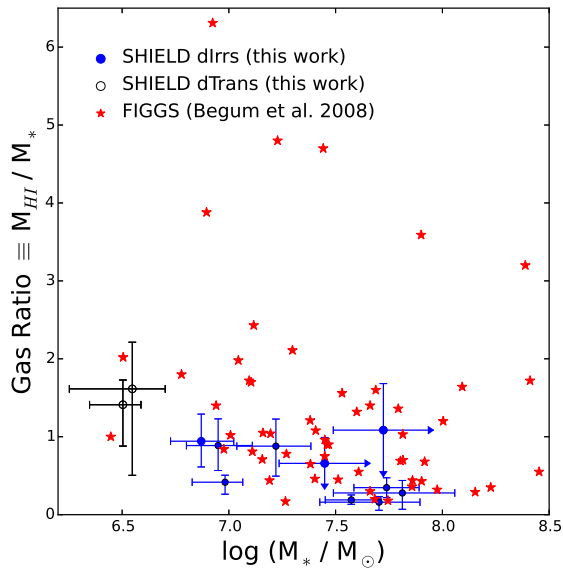}
\caption{An expanded view of Figure~\ref{fig:gas_frac} with results from the FIGGS sample over-plotted. From \citet{Begum2008}, uncertainties on $M_{HI}$ are of order 10\%, while uncertainties on the stellar masses, which are estimated from mass-to-light ratios for the sample, are not well-quantified. The SHIELD sample overlaps with the FIGGS galaxies that have both lower \HI\ masses $\textit{and}$ lower stellar masses.}
\label{fig:gas_frac_figgs}
\end{figure}

Figure~\ref{fig:gas_frac} presents a comparison of the fraction of \HI\ gas mass to stellar mass (M$_{HI}  /  $M$_*$) as a function of stellar mass. These gas fractions  were calculated using the stellar masses derived from the lifetime SFRs and the \HI\ masses based on \HI\ flux measurements from \citet{Cannon2011}. The \HI\ masses and gas fractions are listed in Table~\ref{tab:sf}. The gas properties show a broad distribution as a function of stellar mass with a general trend that gas fractions decrease at higher stellar masses. The trend and range in gas fraction agrees with those previously noted for low-mass galaxies \citep[e.g.,][]{vanZee1997, Schombert2001}. Note that while the fraction of \HI\ mass relative to stellar mass decreases as a function of stellar mass, more massive galaxies in our sample still have overall higher \HI\ masses than do their less massive counterparts. It is this overall larger reservoir of fuel that helps explain the correlation between increasing SFRs with increasing stellar mass in gas-rich systems seen Figure~\ref{fig:sfrs_angst}. Figure~\ref{fig:gas_frac_figgs} expands Figure~\ref{fig:gas_frac} to include results from the Faint Irregular Galaxies GMRT Survey \citep[FIGGS;][]{Begum2008}. We use the \HI\ masses from \citet{Begum2008} and estimate the stellar masses using the B-band luminosities of the galaxies and assuming a mass-to-light ratio of one. The SHIELD galaxies overlap with the low-mass end of the FIGGS systems (i.e., systems with both lower \HI\ masses $\textit{and}$ lower stellar masses).

\begin{figure}
\epsscale{1.15}
\plotone{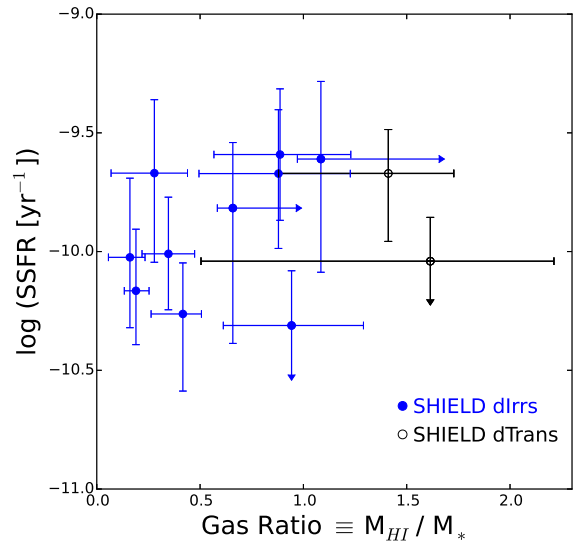}
\caption{Specific SFR as a function of the gas ratios (M$_{HI}$ / M$_*$) for the sample. Similar to Figure~\ref{fig:ssfr_stellar_mass}, there is little correlation between the gas content and star-formation activity once both quantities have been normalized by stellar mass. Despite the scatter, no galaxies show significantly lower levels of SSFRs for a given gas ratio.}
\label{fig:ssfr_gas_ratio}
\end{figure}

\begin{figure}
\epsscale{1.15}
\plotone{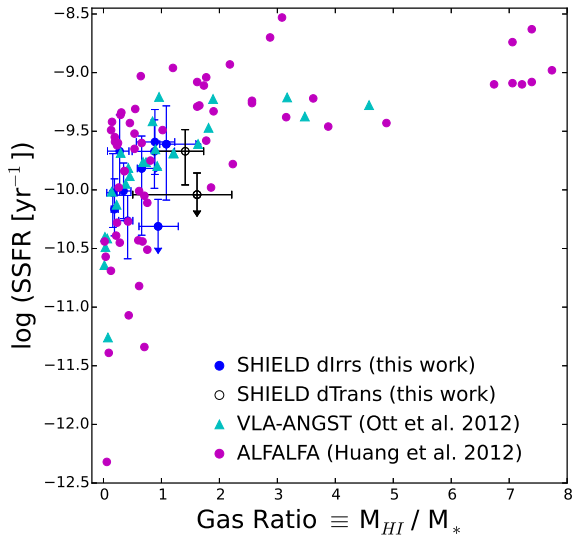}
\caption{An expanded view of Figure~\ref{fig:ssfr_gas_ratio} with results from 2 different surveys over-plotted. Uncertainties for the SSFR measurements are greater than $\pm0.2$ dex for VLA-ANGST and $\pm0.3$ dex for ALFALFA; uncertainties for the gas ratio are not quantified for VLA-ANGST and of order $\pm0.1$ dex for ALFALFA. There is a high degree of dispersion between the SFR and the \HI\ masses when both are normalized by stellar mass. The SHIELD sample overlaps with both surveys at the lower end of their gas ratios. This is not unexpected as the SHIELD sample was selected, in part, based on a criterion of low \HI\ masses.}
\label{fig:gas_ratio_angst}
\end{figure}

Another way of comparing the SFR and gas content is to normalize both parameters by stellar mass. In Figure~\ref{fig:ssfr_gas_ratio}, we present a comparison of the SSFR with the gas fraction (from above, M$_{HI}$ / M$_*$). Consistent with the results presented in Figure~\ref{fig:ssfr_stellar_mass}, the star-formation activity for the SHIELD sample shows a high degree of dispersion.  Figure~\ref{fig:gas_ratio_angst} expands the plot from Figure~\ref{fig:ssfr_gas_ratio} to include the results from the VLA-ANGST survey from \citet{Ott2012} and a sub-sample of the ALFALFA survey with overlapping mass ranges from \citet{Huang2012}. The SHIELD sample lies at the low end of the gas ratio ranges probed by these other surveys. This may be a selection effect as the SHIELD sample was chosen based on low \HI\ masses, and hence may preferentially probe the lower range of gas and stellar mass properties. For gas fractions that overlap with the SHIELD sample, the galaxies from the other surveys span a larger range in SSFRs. This is simply a restatement of our previous finding that the SHIELD galaxies have a narrow range in birthrate parameter. The broadening in SSFR is in contrast with the tight correlation between SSFRs and gas fractions reported for more massive galaxies \citep{Brinchmann2004, Lee2007} and suggests that the stellar mass is not the dominant factor in regulating star formation in low-mass galaxies. Instead, results in this low-mass regime indicate a star-formation process that is dependent upon both the overall gas mass and gas fraction. The star-formation variability in gas-rich low-mass systems is sensitive not only to the overall galaxy mass, but also the star formation history of the galaxy due to intrinsically smaller gas reservoirs. 

Assuming the star formation in each system persists at the recent levels of activity, the gas content in each system can be used to estimate how long a system can continue forming stars. This gas consumption timescale \citep[$\tau_{gas} \equiv$ M$_{gas}$ / SFR, where M$_{gas} \equiv 1.33 \times M_{HI}$ to correct for helium;][]{Roberts1963} assumes a 30\% recycle fraction of stellar mass \citep{Kennicutt1994} and ranges from $2 - 34$ Gyr; individual values are listed in Table~\ref{tab:sf}.  Given the low baryonic content of the galaxies, it is unlikely that their masses have been significantly altered by mergers or gas in-fall. Rather, the longer gas consumption timescales for the SHIELD galaxies indicate that these low-mass galaxies are inefficient at turning their gas content into stars relative to more massive galaxies. 

\section{Conclusions \label{conc} }
We have used optical imaging of resolved stellar populations obtained from the $\textit{HST}$ to measure the star-formation activity of 12 low-mass, gas-rich galaxies in the SHIELD program. The average lifetime SFRs range from $10^{-3.5}$ to $10^{-2.2}$ \msun~yr$^{-1}$. The stellar masses measured from the resolved stellar populations range from $3\times10^6 - 7\times10^7$ \msun, with gas fractions (M$_{HI}$ / M$_*$) between $0.16-1.6$. Overall, the galaxies have recent SFRs comparable to, or slightly higher than, their lifetime SFR averages; none show significantly lower levels of recent star-formation activity compared to historical values. The ratio of recent to lifetime average SFRs fall in a narrow range with a mean value of 1.4 and dispersion of 0.7; this is a much smaller range than reported in other studies of gas-rich, low-mass galaxies \citep[e.g.,][]{Brinchmann2004, Ott2012, Huang2012}. The larger scatter found in these other studies may reflect uncertainties in the measurements or in our comparison (see Section~4), or may be intrinsic to the galaxies. Future work expanding the SHIELD sample to include SFHs of 30 galaxies will provide a larger, homogeneous sample in which to study the birthrate parameter (K.~B.~W.~Mcquinn et al. in preparation). 

We classify two galaxies in the sample, AGC~749241 and AGC~748778, as dTrans galaxies based on their gas content \citep{Cannon2011} and weak or non-detected H$\alpha$ emission \citep{Haurberg2014}. These two systems have star-formation and gas properties consistent with dIrrs both in the SHIELD sample and in previous studies \citep{Begum2008, Bothwell2009, Ott2012, Huang2012}. As both galaxies have measurable recent SFRs based on their resolved stellar populations, the lack of significant H$\alpha$ emission is more likely a result of either stochastic sampling of the upper-end of the IMF or a SFR changing on short timescales, rather than a fundamental difference in galaxy type or properties. 

We used the 3-D distribution of galaxies mapped by \citet{McQuinn2014} around the SHIELD sample to look for correlations between the environments of the galaxies and their recent star-formation properties. Overall, we find no correlation between the distance to the nearest known neighbor and the recent versus lifetime SFR of a system. The more isolated galaxies tend to have higher than average SFRs and the galaxies located near other systems have lower than average SFRs. While we cannot rule out the possibility of an undetected system near the galaxies with higher recent SFRs, no other systems have been found in the ALFALFA data \citep{Haynes2011}, these HST images \citep{McQuinn2014}, nor ancillary VLA data \citep{Cannon2011}. Thus, the recent star-formation activity in these relatively low-density environments appears to be governed by internal evolutionary processes, or the galaxies in question may have experienced a delayed onset to star formation similar to the gas-rich low-mass galaxies Leo~A \citep{Cole2007}, Leo~T \citep{Weisz2012}, and Aquarius \citep{Cole2014}. 

The star-formation activity in these very low-mass galaxies appears to be both inefficient and intrinsically variable. In support of an inefficient star-formation process, not only are the recent and lifetime SFRs low despite being gas-rich, but the oxygen abundances measured from optical spectroscopy are low, indicating an inefficient chemical evolution process \citep{Haurberg2014}. These authors have shown that 2 of the 12 SHIELD galaxies are classified as extremely metal-deficient galaxies \citep[XMD galaxies, e.g.,][]{Kunth2000} with oxygen abundances below 12 $+$ log(O/H) $\leq$ 7.65. In support of a variable, non-deterministic process, we find a high degree of dispersion between the SSFR, stellar mass, and gas fraction of the sample, although we do not report any significantly low values of SSFR for a given stellar mass. This idea of variable or ``flickering'' star-formation activity in low-mass galaxies is further supported by the lack of correlation between systems with higher recent SFRs and the distance to the nearest neighboring system. These results suggest that the star-formation process is governed by the changing internal conditions that are continuously shaped by the star-formation process itself.

Future work on the SHIELD sample will include a spatially resolved study of the \HI\ mass densities and a comparison of the gas properties to the SFRs, infrared emission from Spitzer warm-mission infrared, and ultraviolet emission from GALEX imaging (J.~M.~Cannon et al. in preparation). Additional analysis of the star-formation process from the ground-based H$\alpha$ observations is also forthcoming (N.~C.~Haurberg in preparation). Further, our results will be expanded to include the study of a larger sample of low-mass galaxies identified from the ALFALFA catalog in the SHIELD II survey.

\section{Acknowledgments}
Support for this work was provided by NASA through grant GO-12658 from the Space Telescope Institute, which is operated by Aura, Inc., under NASA contract NAS5-26555. JMC is supported by NSF grant AST-1211683.  Partial support for publication charges was provided by the NRAO. The National Radio Astronomy Observatory is a facility of the National Science Foundation operated under cooperative agreement by Associated Universities, Inc. The authors acknowledge the work of the entire ALFALFA collaboration team in observing, flagging, and extracting the catalogue of galaxies used to identify the SHIELD sample. The ALFALFA team at Cornell is supported by NSF grant AST-1107390 to R.G. and M.P.H. and by a grant to M.P.H. from the Brinson Foundation. This research made use of NASA's Astrophysical Data System and the NASA/IPAC Extragalactic Database (NED) which is operated by the Jet Propulsion Laboratory, California Institute of Technology, under contract with the National Aeronautics and Space Administration. The authors thank the anonymous referee for helpful and constructive comments.

{\it Facilities:} \facility{$\textit{HST}$}

\end{document}